\documentclass[a4paper, 11pt, oneside]{article}
\pdfoutput=1
\usepackage{float}
\usepackage{jheppub}
\usepackage{epstopdf}
\usepackage{feynmp}
\usepackage{bbm}   
\usepackage{amsmath}
\usepackage{graphicx}
\usepackage{slashed}
\usepackage{multirow}
\usepackage{ulem}
\usepackage{graphicx}
\newcommand{\te}{\renewcommand{\arraystretch}{1.4}}
\newcommand{\mc}[3]{\multicolumn{#1}{#2}{#3}}
\newcommand{\nc}{\newcommand}
\nc{\beq}{\begin{equation}}  \nc{\eeq}{\end{equation}}
\nc{\bea}{\begin{eqnarray}}  \nc{\eea}{\end{eqnarray}}
\nc{\baa}{\begin{array}}     \nc{\eaa}{\end{array}}
\nc{\bit}{\begin{itemize}}   \nc{\eit}{\end{itemize}}
\nc{\ben}{\begin{enumerate}} \nc{\een}{\end{enumerate}}
\nc{\bce}{\begin{center}}    \nc{\ece}{\end{center}}
\nc{\bpm}{\begin{pmatrix}}   \nc{\epm}{\end{pmatrix}}
\nc{\bvt}{\begin{verbatim}}  \nc{\evt}{\end{verbatim}}
\nc{\non}{\nonumber} 

\def\nr{\nu_R}
\def\nsm{\nu{\rm SM}}

\def\gev{\;\hbox{GeV}}

\def\hc{\hbox{H.c.}}

\def\vev{vacuum expectation value}

\def\diag{\hbox{\diag}}
\def\lsp{\;\;\;\;}

\def\ui{U(1)}

\def\gcal{{\cal G}}

\def\ocal{{\cal O}}

\def\vcal{{\cal V}}

\def\zBB{{\mathbbm Z}}

\def\ot#1{%
  \mathrel{\vbox{\offinterlineskip\ialign{%
    \hfil##\hfil\cr
    $\scriptscriptstyle(\,\sim\,)$\cr
    \noalign{\kern-.1ex}
    $#1$\cr
}}}}

\def\up#1{^{\left( #1 \right) }}
\def\gdm{\gcal_{\rm dark}}
\def\Box{\partial^2}

\definecolor{mygreen}{rgb}{0,0.5,0}

\begin{document}

\title{Classification of effective operators for interactions\\ 
between the Standard Model and dark matter  }

\author[a]{M. Duch,}
\affiliation[a]{Faculty of Physics, University of Warsaw, Ho\.za 69, 00-681 Warsaw, Poland}
\author[a]{B. Grzadkowski }
\author[b]{J. Wudka}
\affiliation[b]{Department of Physics and Astronomy, UC Riverside,\\
Riverside, CA 92521, USA}
\emailAdd{mateusz.duch@fuw.edu.pl}
\emailAdd{bohdan.grzadkowski@fuw.edu.pl}
\emailAdd{jose.wudka@ucr.edu}
\date{\today}

\abstract{
We construct a basis for effective operators responsible for
interactions between the Standard Model and a dark sector composed of particles 
with spin $ \le1 $. Redundant operators are eliminated using dim-4 equations of motion. 
We consider simple scenarios where the 
dark matter components are stabilized against decay by $\zBB_2$ symmetries.
We determine operators which are loop-generated within an underlying theory and
those that are potentially tree-level generated.
}

\maketitle

\section{Introduction}

Understanding the nature of dark matter (DM) is one of the most pressing current 
issues in astroparticle physics. Of the many hypotheses proposed, one of the most 
fruitful and promising is based on the assumption that DM is composed of one or 
more new elementary particles. This possibility has been extensively studied in 
a variety of specific models, most prominently in realistic supersymmetric models 
that are characterized by predicting that DM is the lightest supersymmetric particle 
whose mass should be in the $O(100 \gev )$ range. 

In this paper we will be interested in constructing a model-independent description of the 
interactions between the dark and SM sectors using an effective Lagrangian approach. 
We will assume the standard sector with one doublet scalar field, but augmented by a
number of right-handed neutrinos $\nr$ to allow for the possibility of Dirac neutrino  masses.
We denote this enlarged model as the $\nsm$. 

Concerning the dark sector we allow  
the dark matter to be multi-component, containing fermions, scalars and 
massive Abelian vectors. Each component is protected against decay 
by a symmetry we denote $ \gdm $, and which we need not to specify at this point, but which can
include discrete or continuous  (local or global) subgroups; 
it also contains the local symmetry related to mass generation of the dark vector.
It is important to note that even though the dark sector may contain
several stable particles, not all have to contribute significantly to the DM relic density
inferred from the WMAP and PLANCK data \cite{Bhattacharya:2013hva}.

Within the scenarios we consider, the dark and $\nsm$ sectors interact through the exchange of
heavy particles whose mass is much larger than the typical momentum transfer in all processes being
considered and their effects decouple at low-energies. In addition, we will assume that the underlying theory 
is weakly coupled and renormalizable. Under these circumstances the DM-$\nsm$ interactions can be described by a series of
effective operators 
\beq
\ocal_{DM-\nsm} =
\ocal_{DM} \ocal_{\nu SM},
\label{oper}
\eeq
where $\ocal_{DM}$ and $\ocal_{\nu SM}$ are composed of fields belonging to the respective 
sectors, and it is assumed that they are invariant under corresponding symmetries.
The  effective Lagrangian consists of a linear combination of terms of this type. The coefficients
are suppressed by appropriate powers of the heavy-physics scale $ \Lambda $ (the power
is determined by the dimension of  $\ocal_{DM-\nsm}$), and contain unknown dimensionless 
couplings that parameterize at low energies all interactions between the standard and dark sectors.
In addition to the hierarchy generated by operator dimensionality it is also useful to note that
some of the operators are necessarily generated by heavy-particle loops, so that their coefficients
are correspondingly suppressed, the remaining operators can be generated at tree level, but whether this is the case
depends on the details of the underlying theory. In this paper we will construct all 
operators $\ocal_{DM-\nsm}$ of dimension $ \le6 $ and determine whether they can be generated at the tree level.

In constructing the effective Lagrangian, we will eliminate operators that
vanish when dim-4 equations of motion are imposed, since they give no contribution to
on-shell matrix elements, both in perturbation theory (to all orders) and
beyond~\cite{Politzer:1980me}; we call such operators redundant.  
A given  type of heavy physics
may generate a basis of operators different from the one listed below; such a basis
may be transformed in the one we use {by applying equations of motion.}

Dim-6 effective operators for interactions between the Standard Model 
and DM have been already present in various contexts in the literature~\cite{Hisano:2014kua}.
However the goal of this paper is to construct a basis of operators~\cite{Duch:2014yma} 
that then could be
consistently adopted to describe different aspects of DM physics.

The paper is organized as follows. In section~\ref{nuSM} we define the model and list 
$\nsm$ operators up to dimension 4.
In section~\ref{dm} we present dark operators needed to construct dim-6 effective operators.
Section~\ref{effLag} contains our main results, i.e. the basis of operators up to dimension 6.
In section~\ref{summary} we summarize of our findings.
Appendix \ref{form} specifies our conventions, while appendix~\ref{vbm} reviews mechanisms for
dark vector boson mass generation.

\section{{\boldmath $\nu SM$} operators}
\label{nuSM}

As mentioned above we will consider the Standard Model of electroweak interactions supplemented by a number of 
right-handed neutrinos $\nu_R$ ($\nu SM$); this model contains the matter fields
collected in table~\ref{sm_mater}. We assume 3 quark families, 3 lepton SU(2) doublets and charged
right-handed lepton singlets, and $ n_\nu$ right-handed neutrinos.
\begin{table*}[h]\centering
\begin{tabular}{|l|cccccc|c|}\hline
 & \mc{6}{c|}{fermions}& scalars\\\hline
field& $l^j_{Lp}$ & \mc{1}{c}{$e_{Rp}$} & \mc{1}{c}{$\nu_{Rk}$} & \mc{1}{c}{$q^{\alpha j}_{Lp}$} 
& \mc{1}{c}{$u^\alpha_{Rp}$} & \mc{1}{c|}{$d^\alpha_{Rp}$} & $\varphi^j$\\ \hline
hypercharge $Y$ & $-\frac{1}{2}$ & \mc{1}{c}{$-1$} & \mc{1}{c}{$0$} & \mc{1}{c}{$\frac{1}{6}$} &
 \mc{1}{c}{$\frac{2}{3}$} & \mc{1}{c|}{$-\frac{1}{3}$} & $\frac{1}{2}$\\ \hline
\end{tabular}
\caption{$\nu SM$ matter field content and their hypercharge quantum numbers. 
Weak isospin, colour and generation indices are denoted by $j=1,2$, $\alpha=1,2,3$ and $p=1,2,3$ respectively. 
We assume the presence of $n_\nu$ right-handed neutrinos, so $k=1,\cdots, n_\nu$.
}
\label{sm_mater}
\end{table*}
We use these fields to construct the gauge-invariant operators $ \ocal_{\nu SM} $ appearing in (\ref{oper}), which we classify according to their canonical dimension (up to dim $ \le 4 $)
and number of Lorentz indices; these operators are collected in table~\ref{tablesm}
(Hermitian conjugation of operators containing fermions are not listed separately but should be included when constructing the effective Lagrangian in order to ensure it is Hermitian).
{It should also be noted that, in the presence of fermionic fields there exists operators where Dirac matrices might appear between the two factors in (\ref{oper}).}
We also use the fact that four-fermion operators can always be rearranged into the form (\ref{oper}) by using Fierz transformations.
All $\nsm$ fields are assumed to be singlets under symmetries stabilizing dark fields.

At this stage, we retain terms that are total derivatives and also we do not apply equations of motion at this point (that will be done when constructing the effective Lagrangian).
\begin{table}[h]
\begin{center}
\te
\small
\begin{tabular}{|c|l|l|l|}
\hline
dim & scalars & vectors & tensors \\
\hline
3/2 & $\nu_{R}$ & - & -\\\hline
2 & $\varphi^\dagger\varphi$ & - & $\overset{(\sim)}{B}_{\mu\nu}$\\\hline
5/2 & $\bar{l}\tilde{\varphi}$ & $\partial_\mu \nu_{R}$ & -\\\hline
3 & $\nu^T_{R} C \nu_{R}$ & $\bar{\psi}\gamma_\mu \psi$, $i\varphi^\dagger\overleftrightarrow{D}_\mu\varphi$ 
& $\nu^T_{R} C \sigma_{\mu\nu} \nu_{R}$, $\partial_\rho\overset{(\sim)}{B}_{\mu\nu}$\\
& & $\partial_\mu(\varphi^\dagger\varphi)$, $\partial^\mu B_{\mu\nu}$ & \\\hline
7/2 & $\varphi^\dagger\varphi\nu_{R}$, $\Box \nu_{R}$ 
& $\bar{l}\partial_\mu\tilde{\varphi}$, $(D_\mu \bar{l})\tilde{\varphi}$ 
& $\partial_\mu\partial_\nu \nu_{R}$, $\nu_{R} \overset{(\sim)}{B}_{\mu\nu}$\\\hline
4 & $(D_{\mu}\varphi)^\dagger D^\mu\varphi$, $\varphi^4$, $\bar\psi \slashed{D}\psi $, & 
  $ \nu^T_{R}  C \partial_\mu \nu_{R} $,  $\nu^T_{R} C  \gamma_{\mu}\slashed{\partial} \nu_{R}$, &
  $\overset{(\sim)}{X}_{\mu\rho}{X_\nu}^\rho$, $\Box\overset{(\sim)}{B}_{\mu\nu}$,  
  $\varphi^\dagger \overset{(\sim)}{W}_{\mu\nu}\varphi$, $\varphi^\dagger \overset{(\sim)}{B}_{\mu\nu}\varphi $,\\
& $\bar{l}\nu_{R}\tilde{\varphi}$, $\bar{l}e{\varphi}$, {$\bar{q}u\tilde{\varphi}$}, $\bar{q}d\varphi$,
& $\nu^T_{R} C \gamma^\mu l \varepsilon \varphi$ &
  $\partial_\mu\partial_\nu (\varphi^\dagger\varphi)$, $ \partial_\mu(i\varphi^\dagger \overleftrightarrow{D}_\nu \varphi)$,
  $(D_\mu\varphi)^\dagger D_\nu \varphi$,
\\
& $\overset{(\sim)}{X}_{\mu\nu}X^{\mu\nu}$, $(D_\mu \bar{\psi} )\gamma^\mu \psi $,   
& &
$\bar{\psi}  D_\mu\gamma_\nu\psi $, $\partial_\mu(\bar{\psi} \gamma_\nu\psi )$, 
$\partial_\mu \partial^\rho B_{\rho\nu}$\\
& $\varphi^\dagger D_\mu D^\mu \varphi$, $(D_\mu D^\mu \varphi)^\dagger \varphi$ & 
&$\bar{l}\sigma^{\mu\nu} \nu_{R} \tilde{\varphi}$, $\bar{l}\sigma^{\mu\nu} e\varphi$ ,
 {$\bar{q}\sigma^{\mu\nu} u \tilde{\varphi}$}, {$\bar{q}\sigma^{\mu\nu} d\varphi$} 
\\\hline
\end{tabular}
\caption{$\nu SM$ operators that are singlets of $SU(3)_C\times SU(2)_L\times U(1)_Y$ in different Lorentz group representations. 
$X_{\mu\nu}$ stands for $B_{\mu\nu}$, $W^I_{\mu\nu}$ or $G^A_{\mu\nu}$, $\psi\in \{l,\nu_R,e,q,u,d\}$.
This list includes operators that are total derivatives, equations of motion were not adopted at this stage.
}
\end{center}
\label{tablesm}
\end{table}

\section{Dark operators}
\label{dm}

In this section we construct the list of operators~\footnote{We omit Lorentz vectors 
and symmetric tensors of dim-4, because the $\nu SM$ does not contain corresponding 
operators of dim~2, so they would be irrelevant while looking for effective operators  
$\nu SM \times DM$ up to dim~6.} up to dim~4, that consist of DM fields: a real scalar $\Phi$, 
left and right chiral fermions $\Psi_L,\Psi_R$ and an Abelian vector field $V_\mu$. 
For simplicity, we assume that the symmetry stabilizing the dark fields is of the form
\beq
\gdm= (\zBB_2)_\Phi \times (\zBB_2)_{\Psi_R}  \times (\zBB_2)_{\Psi_L} \times (\zBB_2)_V
\label{mgdm}
\eeq
where 
the dark scalars $\Phi$ are odd with respect to the first factor and even 
with respect to the others, the $ \Psi_{R(L)}$
are odd with respect to the second (third) factor and even with respects to the others, 
and similarly for the $V_\mu$. {We also introduce a
set of right-handed fermions $N_{R\,l}$ ($l=1,\cdots, n_N$) that transform 
in the same way as $ \Phi $ under $\gdm$. 
As we will show shortly, their presence allows for Yukawa interactions involving
DM and $ \nr$, which might be relevant for DM phenomenology.
As a consequence of these assumptions the lightest particle in each of these
dark sectors ($\Phi$ and $N_R$, $\Psi_L$, $\Psi_R$, $V_\mu$) is stable separately 
and the effective Lagrangian will not contain terms having odd number of fields from any sector.}

In appendix~\ref{vbm} we review two procedures for generating the mass of 
Abelian vector bosons $V_\mu$: the Stuckelberg and Higgs mechanisms.
The dark sector operators for both mechanisms are collected in table~\ref{tabledm}.
Within the Stuckelberg  approach, the $U(1) $ gauge invariance requires that the $V_\mu$ appears
only through the operator $ \vcal_\mu $ or the field strength $ V_{\mu\nu}$; for
the Higgs approach $V_\mu$ appears only within the covariant derivatives of the complex
scalar field $ \phi$ (see appendix~\ref{vbm}).
It should be stressed that $\phi$ is not a dark field 
(as it is explained in the appendix $|\phi|$ belongs to the heavy sector), 
nevertheless we retain $\phi$ in the table to ensure manifest gauge invariance. 
Note that all operators contained in table~\ref{tabledm} are neutral under $\gdm$.
Since all the dark fields are assumed to be singlets under $\nu SM$ gauge
symmetries, so are the operators contained in the table.

\begin{table}[h]
\begin{center}
\te
\small
\begin{tabular}{|c|l|l|l|}
\hline
dim & no space-time indices & one space-time index $\mu$ & more space-time indices $\mu,\nu,\rho$\\	
\hline 
2 & $\Phi^2$ & - & - \\
\hline
5/2 & $\Phi N_R$ & -  & - \\
\hline
3 & $\Psi^T C \Psi$ 
& $\Phi\partial_\mu \Phi$, $\bar\Psi \gamma_\mu\Psi$, $ {\color{blue}\phi^* D_\mu \phi+H.c}$
& $\Psi^T C \sigma_{\mu\nu} \Psi $ \\
\hline
7/2 & - &$\Phi \partial_\mu N_R$, $N_R \partial_\mu \Phi$ & - \\
\hline
4 
& 
$\bar{\Psi}\slashed{\partial}\Psi$, $(\partial_\mu \bar{\Psi})\gamma^\mu \Psi$,
& do not contribute to $ \ocal_{\nsm}$
& $\partial_\mu(\bar{\Psi}\gamma_\nu\Psi)$, $ \bar{\Psi}\partial_\mu\gamma_\nu\Psi$,
\\
& $\Phi^4$, $\partial_\mu\Phi\partial^\mu\Phi$, $\Phi \Box \Phi$,   
& of dim $\le6$.
& ${\color{blue}(D_\mu \phi)^* D_\nu \phi + H.c}$
\\
& 
$\overset{(\sim)}{V}_{\mu\nu} V^{\mu\nu}$, {\color{red}$\mathcal{V}_\mu\mathcal{V}^\mu$}, 
&
&  
\\
& ${\color{blue}(D_\mu\phi)^*D^\mu\phi}$ ,  ${\color{blue}\phi^*D_\mu D^\mu \phi+H.c}$
&
&  \\
\hline
\end{tabular}
\caption{DM operators built of $\Phi$, {$\Psi\in\{\Psi_L, \Psi_R, N_R$\}} and 
$V_\mu $ symmetric under (\ref{mgdm});
note the presence of operators containing $N_R$ and $ \Phi $, allowed by the assumption of
their both being odd under $(\zBB_2)_\Phi $.
For the sake of dark-sector gauge invariance 
the table contains also the vector field $\mathcal{V}_{\mu}$ defined in (\ref{stuc_eq}) 
(highlighted in red) and {covariant derivatives of} the scalar field $\phi$ (blue). They are relevant when the vector 
boson mass is generated through the Stuckelberg and Higgs mechanisms, respectively.
Vector operators and symmetric tensors of dim-4 are not listed, because we are interested 
in the $\nu SM\times DM$ operators up to dim-6 and the $\nu SM$ does not contain vector 
operators nor symmetric tensors with dimension less or equal 2.
}
\label{tabledm}
\end{center}
\end{table}

\section{The effective Lagrangian}
\label{effLag}

In the scenario being considered, the full theory contains not only the dark and standard sectors, 
but also a heavy sector responsible for 
generating effective operators, the theory is assumed to be weakly coupled, and the scale of heavy physics 
$ \Lambda $ is assumed to be substantially larger than the electroweak scale $ v \simeq 246\gev$. At energies significantly below $ \Lambda $
the dynamical content of such a theory is well described by an effective Lagrangian obtained by ``integrating out'' the heavy degrees of freedom, and which takes the form
\begin{equation}
 \mathcal{L}=\mathcal{L}^{(4)}+\frac{1}{\Lambda}\sum_{k}C\up5_k O_k^{(5)}+\frac{1}{\Lambda^2}\sum_{k}C\up6_k O_k^{(6)} + \cdots,
\end{equation}
where each term $O_k\up n$ is of the form (\ref{oper}) and is multiplied by an unknown 
dimensionless (Wilson) coefficient $C_k\up n$.
It should be stressed that the right-handed neutrinos $\nu_R$ belong to the light sector, 
so the
scale of their mass must be $ \ll \Lambda $; in the following we will assume that it 
is at most of the order of $v$.

The effective Lagrangian is a useful tool for parametrizing the low-energy effects of a theory 
in a consistent and controlled way. In the case where the low-energy theory is described by
the SM the full set of such operators up to dim-6 
was constructed in~\cite{Buchmuller:1985jz} and then refined and systematized in 
\cite{Grzadkowski:2010es}. 

As we have already mentioned, we allow in the dark sector for the presence of $N_R$, right-handed
fermions which transform the same way as $\Phi$ under $\gdm$. 
As we will shortly demonstrate, this choice has important consequences since then 
Yukawa-type interactions between the right-handed neutrinos $\nr$, the $N_R$ and $\Phi$ 
(the last two are odd under $(\zBB_2)_\Phi$) are allowed~\cite{Grzadkowski:2009mj}. 
This represents a new renormalizable portal coupling between the $\nsm$  and the dark sector.

Note that since the $N_R$ are odd under $\gdm$, they cannot 
mix with the $ \nu_L$ to generate a Dirac mass term; the $ \nu_R$, however,
can.

The terms of dimension $\le4$ in the effective Lagrangian, $\mathcal{L}^{(4)}$, consists of 3 parts
\begin{equation}
 \mathcal{L}^{(4)}=\mathcal{L}^{(4)}_{\nu SM}+\mathcal{L}^{(4)}_{DM}+\mathcal{L}^{(4)}_{\nu SM\times DM}.
\label{lag}
\end{equation}

The first part is the Standard Model Lagrangian with right-handed neutrinos
\bea
\mathcal{L}^{(4)}_{\nu SM} & = &-\frac{1}{4}G^A_{\mu\nu}G^{A\mu\nu} -\frac{1}{4}W^I_{\mu\nu}W^{I\mu\nu} -\frac{1}{4}B_{\mu\nu}B^{\mu\nu}
+(D_\mu \varphi)^\dagger(D^\mu \varphi) + m^2 \varphi^\dagger\varphi - \frac{1}{2}\lambda(\varphi^\dagger\varphi)^2\non\\
&&+i(\bar{l}\slashed{D} l + \bar{\nu}_R\slashed{\partial} \nu_R+\bar{e}\slashed{D}e+\bar{q}\slashed{D} q+\bar{u}\slashed{D} u+\bar{d}\slashed{D} d)\label{SMLag} \\
&&-(\bar{l}\Gamma_\nu \nu_R\tilde{\varphi}+ \bar{l}\Gamma_e e\varphi + \bar{q}\Gamma_u u \tilde{\varphi} + \bar{q}\Gamma_d d\varphi + \hc )-
\frac12 (\nu^T_R  C m_\nu \nu_R  +  \hc ),\non
\eea
where $\Gamma_{e,u,d}$ are $3\times 3$ matrices, $\Gamma_\nu$ is a $3\times n_\nu$ matrix and Majorana mass $m_\nu$ 
is $n_\nu \times n_\nu$ matrix.  
Note that we can always choose a field basis such that $m_\nu$ is diagonal.

The second term in (\ref{lag}) contains only dark fields~\footnote{Note that Dirac mass terms $\bar{\psi}_L \psi_R$ 
and $\bar{\psi}_L N_R$ are forbidden by $\gdm$ symmetries.}
\begin{equation}
\begin{split}
 \mathcal{L}^{(4)}_{DM} =  & \frac{1}{2} \partial_\mu \Phi \partial^\mu \Phi -\frac{1}{2} m^2_\Phi \Phi^2 - 
\frac{1}{4}\kappa \Phi^4 -\frac{1}{4} V^{\mu\nu} V_{\mu\nu} + \frac{1}{2} m_V^2 V_\mu V^\mu \\
 & + i (\bar{\Psi}_L \slashed{\partial} \Psi_L + \bar{\Psi}_R\slashed{\partial}\Psi_R)
- \frac12 ( \Psi^T_L C m_L \Psi_L + \Psi^T_R C m_R \Psi_R + \hc ),\\
 & + i \bar{N}_R\slashed{\partial}N_R - \frac12 (N^T_R C m_N N_R + \hc ),
\label{DMLag}
\end{split}
\end{equation}
where $m^2_\Phi>0$ in order to preserve $ (\zBB_2)_\Phi$. $m_N$ is a $n_N\times n_N$ matrix that, as in the case of the $ \nu_R$, can be assumed to be diagonal; 
possible mechanisms for vector-boson-mass generation are reviewed in appendix~\ref{vbm}.

The last term in (\ref{lag}), responsible for dim-4 interactions between $\nsm$ and DM, reads
\begin{equation}
  \mathcal{L}^{(4)}_{\nu SM\times DM}=g_\varphi \varphi^\dagger\varphi \Phi^2 + 
	(\nu^T_R C Y_\Phi N_R + \hc)\Phi, 
	\label{Lag_int}
\end{equation}
where $Y_\Phi$ is a $n_\nu\times n_N$ matrix. 

Except of the standard Higgs portal, 
$\varphi^\dagger\varphi \Phi^2$, following~\cite{Grzadkowski:2009mj}, we have added above 
a possible Yukawa interactions between $\nu_R$, $N_R$ and the dark scalar $\Phi$. 
Here we have also assumed that the dark fields can carry only 
their own $\zBB_2$ quantum numbers, so a given dark field can not transform non-trivially under 
$\zBB_2$ that stabilizes a different dark sector component, this eliminates some operators
that otherwise would be present, e.g. $\nu^T_R C \Psi_R \Phi$. {Note that the kinetic mixing between 
the $U(1)_Y$ and the additional $U(1)$ corresponding to the dark vector boson $V_\mu$ is forbidden by the
stabilizing symmetry $\gdm$. }
The Lagrangian $\mathcal{L}^{(4)}_{\nu SM\times DM}$ contains all possible 
renormalizable interactions between the $\nu SM$ and DM that are allowed within the assumptions specified above.

{We wish to make a comment concerning stability of fermions that appear in our scenario
and which are neutral under SM gauge symmetries. 
We assumed that there are $n_\nu$ right-handed neutrinos $\nr$.
In general $\nr$'s decay by standard Yukawa interactions, 
however the lightest of them is stable. Besides $\nr$'s there are $n_N$ of $N_R$'s, $\Psi_R$ and $\Psi_L$.
Among then only $\Psi_R$ and $\Psi_L$ are guarantied to be stable 
(by the virtue of $(\zBB_2)_{\Psi_R}  \times (\zBB_2)_{\Psi_L}$). 
Since both $N_R$'s and $\Phi$ are odd under $\gdm$, therefore the lightest of them is stable,
in other words $N_R$'s might be unstable.}

Equations of motion for $\nu SM$ fields derived from the $\mathcal{L}^{(4)}$ for $\nu SM$ 
fields are the same as in the SM with {two}  exceptions. 
Equations for the Higgs doublet ($\varphi$) and right-handed neutrinos ($\nu_R$) contain terms, 
that originate from interactions {present} in $\mathcal{L}^{(4)}_{\nu SM\times DM}$. 
The complete list of equations of motion for $DM$ and $SM\nu_R$ fields is:
\begin{equation}
 \begin{split}
&(D_\mu D^\mu\varphi)^j=m^2\varphi^j - 
\lambda (\varphi^\dagger\varphi)\varphi^j-\varepsilon_{jk}\bar{l}^k\Gamma_\nu \nu_R 
-\bar{e}\Gamma^\dagger_e l^j -
\varepsilon_{jk}\bar{q}^k\Gamma_u u - \bar{d}\Gamma^\dagger_d q^j + g_\varphi\varphi \Phi^2,\\
&(D^\rho G_{\rho\mu})^A=g_s(\bar{q}\gamma_\mu T^Aq + \bar{u}\gamma_\mu T^A u + \bar{d}\gamma_\mu T^A d),\\
&(D^\rho W_{\rho\mu})^I=\frac{g}{2}(\varphi^\dagger i \overleftrightarrow{D}^I_\mu\varphi+ \bar{l}\gamma_\mu \tau^I l 
+\bar{q}\gamma_\mu \tau^I q),\\
&\partial^\rho B_{\rho\mu} = g'Y_\varphi \varphi^\dagger i \overleftrightarrow{D}_\mu\varphi 
+ g' \sum_{\psi\in\{l,e,q,u,d\}} Y_\psi\bar{\psi}\gamma_\mu\psi
\end{split}
\label{eof_v}
\end{equation}
\begin{equation}
\begin{split}
&i\slashed{D}l=\Gamma_\nu \nu_R \tilde{\varphi} + \Gamma_e e \varphi,\\
&i\slashed{\partial}\nu_R=\Gamma^\dagger_\nu \tilde{\varphi}^\dagger l + m_\nu \nu^C_R - Y^\dagger_\Phi N^C_R \Phi \\\
&i\slashed{\partial}N_R= m_N N^C_R - Y^\dagger_\Phi \nu^C_R \Phi \\\
&i\slashed{D}e=\Gamma^\dagger_e\varphi^\dagger l,\\
&i\slashed{D}q=\Gamma_u u \tilde{\varphi} + \Gamma_d d \varphi,\\
&i\slashed{D}u=\Gamma^\dagger_u \tilde{\varphi}^\dagger q,\\
&i\slashed{D}d=\Gamma^\dagger_d \varphi^\dagger q\\
\end{split}
\label{eof_f}
\end{equation}
\begin{equation}
\begin{split}
&\partial_\mu \partial^\mu \Phi = -m_\Phi \Phi - \kappa \Phi^3 + 2 g_\varphi \Phi \varphi^\dagger \varphi 
+ \nu^T_R C Y_\Phi N_R + \bar{N}_R C Y^\dagger_\Phi \bar{\nu}^T_R,\\
&\partial^\mu V_{\mu\nu} = - m_V V_\nu,\;\;\;\;\;\;\;\partial_\mu V^{\mu}=0,\\
&i \slashed{\partial} \Psi_{L,R} = m_{L,R} \Psi^c_{L,R}.
\end{split}
\label{eof_dm}
\end{equation}
Many operators of the form $\nu SM\times DM$  are redundant through the application of the 
equations of motion~\cite{Politzer:1980me}, and should be omitted from  the  basis. 
Below we provide an illustration of this process of elimination; the notation we use is the following.
If an operator includes LHS of one of the above equations, then
it can be written as a sum of the operators that consists of the RHS of that equation and an operator, 
denoted by $\boxed{EOM}$, which vanishes due to that equation of motion. 
The purpose is to express a given operator as a linear combination of other operators, 
total derivatives $\boxed{TD}$ and $\boxed{EOM}$. 
Such operators are redundant in effective Lagrangian. Operators vanishing due to the Bianchi 
identity are denoted by $\boxed{BI}$. Using tab. \ref{tablesm} and \ref{tabledm} and these rules one can 
construct an irreducible basis of $\nu SM\times DM$ operators up to dim~6. 

We provide two examples of how the equations of motion can be used to eliminate some operators.
First we show that $\bar{\psi}\gamma_\mu\psi \partial^\mu  (\Phi^2)$ ($\psi\in\{l,\nu_R,e,q,u,d\}$)
is redundant. After integrating by parts and applying equation of motion (\ref{eof_f})
we obtain the following
\beq
\bar{\psi}\gamma_\mu\psi \partial^\mu  (\Phi^2) = \boxed{TD} - \partial_\mu  (\bar{\psi}\gamma_\mu\psi)\Phi^2 = 
\boxed{TD} - (\bar{\psi} \slashed{D}\psi + h. c. )\Phi^2 = 
\boxed{TD} + \boxed{EOM} +  O^{(4)}_{\nu SM} \times \Phi^2,
\eeq
where $O^{(4)}_{\nu SM} \times \Phi^2$ denotes operators made as a product of some operator
belonging to $\mathcal{L}^{(4)}_{\nu SM}$ and $\Phi^2$. If all operators $O^{(4)}_{\nu SM} \times \Phi^2$
are included in our list then there is no need to have $\bar{\psi}\gamma_\mu\psi \partial^\mu  (\Phi^2)$ as well.

A bit more involved algebra is needed to show redundancy of 
$B^{\mu\nu}\bar{\Psi} { \gamma_\mu \partial_\nu} \Psi$
\bea
  B^{\mu\nu}\bar{\Psi} { \gamma_\mu \partial_\nu} \Psi & =	&\frac{1}{2} B^{\mu\nu}\bar{\Psi}
  (\gamma_\mu \gamma_\nu \slashed{\partial} + \gamma_\mu \slashed{\partial} \gamma_\nu)\Psi =  
	\frac{1}{2} B^{\mu\nu}\bar{\Psi}
  (\gamma_\mu \gamma_\nu \slashed{\partial}-\slashed{\partial} \gamma_\mu \gamma_\nu)\Psi+
 B^{\mu\nu}\bar{\Psi}\gamma_\nu \partial_\mu \Psi=\non\\
 & &\frac{1}{4} B^{\mu\nu}\bar{\Psi}
  (\gamma_\mu \gamma_\nu \slashed{\partial} - \slashed{\partial} \gamma_\mu \gamma_\nu)\Psi =
 \frac{1}{4} B^{\mu\nu}\bar{\Psi}\gamma_\mu \gamma_\nu \slashed{\partial}\Psi + 
\frac{1}{4}\bar{\Psi}\overleftarrow{\slashed{\partial}}\gamma_\mu \gamma_\nu \Psi B^{\mu\nu} + \label{longest} \\
& &+\frac{1}{4}\bar{\Psi}\gamma_\rho\gamma_\mu\gamma_\nu\Psi \partial^\rho B^{\mu\nu} + \boxed{TD} =
\boxed{EOM} + \boxed{TD} + \boxed{BI} \non\\ 
& &+\frac{1}{4} (B_{\mu\nu} \Psi^T C \sigma^{\mu\nu} \Psi+ H.c.)
+\frac{ig'}{4}\varphi^\dagger\overleftrightarrow{D}_\mu\varphi\bar\Psi \gamma^\mu \Psi 
+ \frac{g'}{2} \sum_{\psi\in\{l,e,q,u,d\}} Y_\psi \bar{\psi} \gamma_\mu \psi \bar{\Psi} \gamma^\mu \Psi.\non
\eea
where we used
\bea
 \bar{\Psi}\gamma_\rho\gamma_\mu\gamma_\nu\Psi \partial^\rho B^{\mu\nu} &=&
2\bar{\Psi}\gamma^\nu \Psi \partial^\rho B_{\rho\nu}
-\bar{\Psi}i\varepsilon_{\rho\mu\nu\sigma}\gamma^\sigma\gamma_5 \Psi \partial^\rho B^{\mu\nu} = 
\label{gggB}\\
&=&\boxed{EOM} + \boxed{BI} + 
ig'\varphi^\dagger\overleftrightarrow{D}_\mu\varphi\bar\Psi \gamma^\mu \Psi 
+ 2g' \sum_{\psi\in\{l,e,q,u,d\}} Y_\psi \bar{\psi} \gamma_\mu \psi \bar{\Psi} \gamma^\mu \Psi.\non
\eea
Again, if operators $i\varphi^\dagger\overleftrightarrow{D}_\mu\varphi\bar\Psi \gamma^\mu \Psi$ and 
$\bar{\psi}_p \gamma_\mu \psi_q \bar{\Psi} \gamma^\mu \Psi$ were present 
then  $B^{\mu\nu}\bar{\Psi}  \gamma_\mu \partial_\nu \Psi$ should be omitted from the
operator basis. 
Similar arguments apply if $B^{\mu\nu}$ is replaced by $\tilde{B}^{\mu\nu}$. 

Before proceeding to 
the final table collecting all the effective operators we discuss the
mechanisms of dark vector boson mass generation, as they are relevant for the final output.

\subsection{The Stuckelberg mechanism}

In this scenario the Lagrangian (\ref{stuckel}) is invariant under $\zBB_2$ symmetry, with
both $\sigma$ and $V_\mu$ being odd. The equations of motion read
\begin{equation}
\begin{split}
 &\partial_\mu(\partial^\mu\sigma - m_V V^\mu)\equiv\partial_\mu  \mathcal{V}^\mu = 0\\
 &\partial_\mu V^{\mu\nu} =  -m^2_V V^\nu + m_V \partial^\nu \sigma = m_V \mathcal{V}^\nu.
\label{stuc_eq}
\end{split}
\end{equation}
The Stuckelberg Lagrangian and equations of motion reduce to a part of the Lagrangian (\ref{DMLag}) 
and equations (\ref{eof_dm}) when the $ \sigma =0 $ (unitary) gauge is adopted.

It is worth, at this point, to discuss in some detail the operator composed of two dark vector fields 
and two Higgs boson doublets: 
$V_\mu V^\mu \varphi^\dagger\varphi$, the effects of which have been investigated
in the literature e.g. in \cite{Lebedev:2011iq}.
Because of gauge invariance this operator can only be generated by 
$\mathcal{V}_\mu\mathcal{V}^\mu\varphi^\dagger\varphi$ of mass dimension 6. It should be noticed that
in the $\sigma=0$ gauge 
\begin{equation}
  \frac{1}{\Lambda^2} \mathcal{V}_\mu\mathcal{V}^\mu \varphi^\dagger\varphi\rightarrow \frac{m_V^2}{\Lambda^2} V_\mu V^\mu  \varphi^\dagger\varphi,
\end{equation}
therefore there appears an unavoidable suppression factor $m_V^2/\Lambda^2$ even though formally 
$V_\mu V^\mu  \varphi^\dagger\varphi$ is dim-4 operator.
Note that higher dimensional operators of that sort would be suppressed by higher powers of $\Lambda$.

\subsection{The Higgs mechanism}

Another method to generate vector mass is the Higgs mechanism (\ref{higgs}).
In this case $ (\zBB_2)_V$ corresponds to charge conjugation:
writing $ \phi = \rho \exp(i \theta) $, $V_\mu$ and $ \theta $ are odd
while $ \rho$ is even. Since we require $V_\mu$ to be stable,  
the Lagrangian must be invariant under $(\zBB_2)_V$ and this symmetry
must remain unbroken. This is  indeed what happens since only $ \rho $ {develops}  a \vev, 
for details see appendix~\ref{vbm}. The vector-boson mass $m_V$ is of order of the Higgs field \vev\ $v$ 
provided the gauge coupling constant $g\sim v/\Lambda$ and $\langle\rho\rangle \sim \Lambda$.
The remaining physical scalar has  a mass $ \sim \Lambda$ and decouples from the low-energy theory.

Let us again focus on the $V_\mu V^\mu \varphi^\dagger\varphi$ operator. 
It is easy to see that it can only be generated by
$|D_\mu \phi|^2( \varphi^\dagger \varphi)$. Using now the above values for $g$ and $ \langle\rho\rangle$ we find
\begin{equation}
  \frac{1}{\Lambda^2}(D_\mu \phi)^*D^\mu\phi \varphi^\dagger \varphi
 \rightarrow \frac{(gv)^2}{\Lambda^2}  V_\mu V^\mu \varphi^\dagger\varphi = \frac{m_V^2}{\Lambda^2}  V_\mu V^\mu \varphi^\dagger\varphi,
\end{equation}
so we observe that this coupling is also suppressed in the Higgs scenario.

Few other comments are here in order. First, note that the
operator $\phi^*D_\mu D^\mu \phi$ is not invariant under $(\zBB_2)_V$. 
In order to make it invariant we have to add its conjugate:
\begin{equation}
 (\phi^*D_\mu D^\mu \phi+(D_\mu D^\mu \phi)^*\phi)\varphi^\dagger\varphi.
\end{equation}
Similarly $(D_\mu\phi)^*D_\nu\phi + (D_\nu\phi)^*D_\mu\phi$ is invariant under $(\zBB_2)_V$. 
Note that this operator is symmetric in its Lorentz indices and vanishes after 
contraction with ${B}_{\mu\nu}$.

\subsection{{\boldmath $\nu SM \times DM$} operators}

The resulting effective operators obtained via (\ref{oper}) using tables \ref{tablesm} and \ref{tabledm}
are contained in table~\ref{tablesmdm}.
\begin{table*}[ht]
\centering
\begin{tabular}{|c|c|c|c|c|c|} 
\hline
& $1$ & $\Lambda^{-1}$ & \mc{3}{|c|}{$\Lambda^{-2}$} \\ 
\hline
& & & $\Phi$,{$N_R$} & $\Psi$ & $V_{\mu}$ \\
\hline
\textsc{Tree}:& $\varphi^\dagger\varphi \Phi^2$ &  $\varphi^\dagger\varphi \Psi^T C \Psi$ &  
$\varphi^\dagger\varphi \partial_\mu \Phi \partial^\mu \Phi$
& $\nu^T_{R} C \nu_{R} \Psi^T C \Psi$ & $m_V^2\varphi^\dagger\varphi V_\mu V^\mu$\\
& $\nu^T_{R} C N_{R}\Phi$ & {$\bar{l} \tilde{\varphi} N_R  \Phi$} & $\varphi^\dagger\varphi \Phi^4$
& $\nu^T_{R} C \sigma^{\mu\nu} \nu_{R} \Psi^T C \sigma_{\mu\nu} \Psi$ & \\
&& $\nu^T_R C \nu_R \Phi^2$ & $(\varphi^\dagger\varphi)^2\Phi^2$
& $\nu^T_{R} C \nu_{R} \bar{\Psi} C \bar{\Psi}^T$ &\\
&& & $\bar{l}\nu_{R}\tilde{\varphi}\Phi^2$ 
& $\nu^T_{R} C \sigma^{\mu\nu} \nu_{R} \bar{\Psi} \sigma_{\mu\nu} C \bar{\Psi}^T$ &\\
&& &  $\bar{l}e{\varphi}\Phi^2$& $\bar{\psi}_p \gamma_\mu \psi_q \bar\Psi \gamma^\mu \Psi$& \\
&& &  $\bar{q}u\tilde{\varphi}\Phi^2$ & $ i\varphi^\dagger\overleftrightarrow{D}_\mu\varphi\bar\Psi \gamma^\mu \Psi$&\\
&& &  $\bar{q}d{\varphi}\Phi^2$& &\\
&& &  $\nu^T_{R}CN_{R}\Phi^3$& &\\
&& &  $\varphi^\dagger\varphi\nu^T_{R}CN_{R}\Phi$& &\\
&& &  $N^T_{R} C \gamma^\mu l \varepsilon \partial_\mu \varphi \Phi$& &\\\hline
\textsc{Loop:}&& $\overset{(\sim)}{B}_{\mu\nu} \Psi^T C \sigma^{\mu\nu} \Psi$& 
$\nu^T_{R}C\partial_\mu N_{R}\partial^\mu\Phi$ & &
$\varphi^\dagger\varphi\overset{(\sim)}{V}_{\mu\nu}V^{\mu\nu}$\\
&& & $\nu^T_{R} C \sigma^{\mu\nu} N_{R}\overset{(\sim)}{B}_{\mu\nu}\Phi$ & & $m_V^2\varphi^\dagger\varphi V_\mu V^\mu$\\
&&&$\overset{(\sim)}{X}_{\mu\nu} X^{\mu\nu} \Phi^2$&&\\
\hline
\end{tabular}
\caption{List of all $\nu SM\times DM$ operators up to dim~6, that are suppressed 
by at most~$\Lambda^{-2}$.
Dark matter sector consists of a real scalar $\Phi$, chiral fermions 
{$\Psi\in\{\Psi_L, \Psi_R, N_R$\}} and vector field $V_\mu$. 
Tree and loop-generated operators are collected in the upper and lower part of the table, 
respectively. Operator $\varphi^\dagger\varphi V_\mu V^\mu$ appears in both categories, 
because within the Higgs mechanism, it can be generated at the tree-level approximation,
while within the Stuckelberg model it requires a loop.
Note that one~entry in the table may refer to various operators, because $(\sim)$ over 
$X_{\mu\nu}$ denotes $X_{\mu\nu}$ or~$\tilde{X}_{\mu\nu}$,
 $X_{\mu\nu}$ stands for $B_{\mu\nu}$, $W^I_{\mu\nu}$ or $G^A_{\mu\nu}$ and 
$\psi\in \{l,\nu_{R},e,q,u,d\}$.
The bosonic operators are all Hermitian. In case of the operators containing fermions, 
$i\varphi^\dagger\protect\overleftrightarrow{D}_\mu\varphi\bar\Psi \gamma^\mu \Psi$  
is Hermitian and conjugation of 
$\bar{\psi}_p \gamma_\mu \psi_q \bar\Psi \gamma^\mu \Psi$ is equivalent to transposition
of the generation indices. 
For the remaining operators Hermitian conjugations are not listed explicitly.
}
\label{tablesmdm}
\end{table*}
There are several comments here in order.

Note that operators denoted by ``TREE" in table~\ref{tablesmdm} are those for which there 
exists a new physics model where they are generated at tree level, but it is not yet possible 
to determine whether this is the case for the situation realized in Nature; a specific model 
may generate those potentially tree generated (PTG) operators at one or higher loops, or may not generate it at all 
because of the details of its particle content and symmetries.

All operators present in table~\ref{tablesmdm} are of the form of (\ref{oper}) 
with ${\cal O}_{\rm \nu SM}$ and ${\cal O}_{\rm DM}$ being separately invariant 
under symmetries of $\nsm$ and $\gdm$, respectively. 
Such operators can be generated at tree-level by the exchange of heavy particles that may or may not be neutral under
the $\nsm$ and dark symmetries (both options always exist, though existing 
data may constrain the properties of non-neutral particles more severely).
For example $\varphi^\dagger\varphi \Psi_L^T C \Psi_L $ can be generated by the exchange 
of a neutral heavy scalar $S$ with couplings $ S |\varphi|^2 $
and   $\Psi_L^T C \Psi_L \, S $; or by a heavy Dirac fermion $F$ 
with the same SM gauge transformation properties as $ \varphi $, odd under $ (\zBB_2)_{\Psi_L}$, 
and with couplings $\bar F \varphi \Psi_L$  and $ \bar F \varphi ( \Psi_L)^C $ 
($C$ denotes the usual charge conjugation operation).
For an illustration, in table~\ref{diag},
we draw generic diagrams (within an underlying theory) that could be responsible for operators contained in 
table~\ref{tablesmdm}. 

It should be noticed that the stabilizing symmetries imply that neither $\Psi_L$ nor $\Psi_R$ can appear
separately in any interaction. Therefore it is sufficient to restrict ourself just to one chirality 
{of $\Psi$, consequently one could drop e.g. $\Psi_R$, such that $\Psi$ in table~\ref{tablesmdm}
would just correspond to $\Psi_L$ or $N_R$}.

%
\begin{table}[t]
\begin{center}
\vspace*{-1cm}
\begin{tabular}{|l|l|l|}\hline
& \multicolumn{2}{|c|}{mediator}\\\hline
operator & neutral & charged\\\hline
$\varphi^\dagger\varphi\Psi^T C \Psi$ & \begin{fmffile}{a1}
\fmfstraight
	        \begin{fmfgraph*}(60,25)
 	            \fmfset{arrow_len}{0.25cm}
 	            \fmfset{arrow_ang}{13}
 	            \fmfset{dash_len}{2mm}
	            \fmfleft{i,j}
	            \fmfright{k,l}
	            \fmf{scalar}{i,v1}
		    \fmf{scalar}{v1,j}
		    \fmf{dashes}{v1,v2}
		    \fmf{fermion}{k,v2}
		    \fmf{fermion}{v2,l}
		    \fmfdot{v1}
	        \end{fmfgraph*}
	    \end{fmffile} &
\begin{fmffile}{a2}
\fmfstraight
	        \begin{fmfgraph*}(60,25)
 	            \fmfset{arrow_len}{0.25cm}
 	            \fmfset{arrow_ang}{13}
 	            \fmfset{dash_len}{2mm}
	            \fmfleft{i,j}
	            \fmfright{k,l}
	            \fmf{scalar}{i,v1}
		    \fmf{fermion}{v1,j}
		    \fmf{fermion}{v2,v1}
		    \fmf{fermion}{k,v2}
		    \fmf{scalar}{v2,l}
	        \end{fmfgraph*}
	    \end{fmffile}
 \\\hline
$\bar{l}\tilde{\varphi}N_R\Phi$ & \begin{fmffile}{z1}
\fmfstraight
\begin{fmfgraph*}(60,25)
 	            \fmfset{arrow_len}{0.25cm}
 	            \fmfset{arrow_ang}{13}
 	            \fmfset{dash_len}{2mm}
	            \fmfleft{i,j}
	            \fmfright{k,l}
	            \fmf{scalar}{i,v1}
		    \fmf{fermion}{v1,j}
		    \fmf{fermion}{v2,v1}
		    \fmf{fermion}{k,v2}
		    \fmf{dashes}{v2,l}
	        \end{fmfgraph*}
	    \end{fmffile}
	        &
\begin{fmffile}{z2}
\fmfstraight
	        \begin{fmfgraph*}(60,25)
 	            \fmfset{arrow_len}{0.25cm}
 	            \fmfset{arrow_ang}{13}
 	            \fmfset{dash_len}{2mm}
	            \fmfleft{i,j}
	            \fmfright{k,l}
	            \fmf{scalar}{i,v1}
		    \fmf{dashes}{v1,j}
		    \fmf{scalar}{v1,v2}
		    \fmf{fermion}{k,v2}
		    \fmf{fermion}{v2,l}
		    \fmfdot{v1}
	        \end{fmfgraph*}
	    \end{fmffile} 
 \\\hline

$\nu^T_R C \nu_R \Phi^2$ & 
\begin{fmffile}{b1}
	        \begin{fmfgraph*}(60,25)
 	            \fmfset{arrow_len}{0.25cm}
 	            \fmfset{arrow_ang}{13}
 	            \fmfset{dash_len}{2mm}
	            \fmfleft{i,j}
	            \fmfright{k,l}
	            \fmf{dashes}{k,v1}
		    \fmf{dashes}{v1,l}
		    \fmf{dashes}{v1,v2}
		    \fmf{fermion}{i,v2}
		    \fmf{fermion}{v2,j}
		    \fmfdot{v1}
	        \end{fmfgraph*}
	    \end{fmffile}
 & 
\begin{fmffile}{b2}
	        \begin{fmfgraph*}(60,25)
 	            \fmfset{arrow_len}{0.25cm}
 	            \fmfset{arrow_ang}{13}
 	            \fmfset{dash_len}{2mm}
	            \fmfleft{i,j}
	            \fmfright{k,l}
	            \fmf{dashes}{k,v1}
		    \fmf{dashes}{v2,j}
		    \fmf{fermion}{v2,v1}
		    \fmf{fermion}{i,v2}
		    \fmf{fermion}{v1,l}
	        \end{fmfgraph*}
	    \end{fmffile}
\\\hline
$\varphi^\dagger\varphi\partial_\mu\Phi\partial^\mu\Phi$ & &
\begin{fmffile}{c1}
	        \begin{fmfgraph*}(60,25)
		    \fmfset{arrow_len}{0.25cm}
 	            \fmfset{arrow_ang}{13}
 	            \fmfset{dash_len}{2mm}
	            \fmfleft{i,j}
	            \fmfright{k,l}
	            \fmf{dashes}{k,v1}
		    \fmf{scalar}{v1,l}
		    \fmf{boson}{v1,v2}
		    \fmf{scalar}{i,v2}
		    \fmf{dashes}{v2,j}
		    \fmfv{lab=$\partial_\mu$,lab.dist=0.05w}{v1}
		    \fmfv{lab=$\partial^\mu$,lab.dist=0.05w}{v2}
	        \end{fmfgraph*}
	    \end{fmffile} 
 \\\hline
 
$\varphi^\dagger\varphi \Phi^4$ &
\begin{fmffile}{d1}
	        \begin{fmfgraph*}(60,25)
 	            \fmfset{arrow_len}{0.25cm}
 	            \fmfset{arrow_ang}{13}
 	            \fmfset{dash_len}{2mm}
	            \fmfbottom{i,l,m}
	            \fmftop{j,k,n}
	            \fmf{scalar}{i,v1}
		    \fmf{scalar}{v1,j}
		    \fmf{dashes}{v1,v2}
		    \fmf{dashes}{v2,v3}
		    \fmf{dashes}{k,v2}
		    \fmf{dashes}{v2,l}
		    \fmf{dashes}{m,v3}
		    \fmf{dashes}{n,v3}
		    \fmfdot{v1,v3}
	        \end{fmfgraph*}
	    \end{fmffile}
 & \begin{fmffile}{d2}
	        \begin{fmfgraph*}(60,25)
 	            \fmfset{arrow_len}{0.25cm}
 	            \fmfset{arrow_ang}{13}
 	            \fmfset{dash_len}{2mm}
	            \fmfbottom{i,l,m}
	            \fmftop{j,k,n}
	            \fmf{scalar}{i,v1}
		    \fmf{dashes}{v1,j}
		    \fmf{scalar}{v1,v2}
		    \fmf{scalar}{v2,v3}
		    \fmf{dashes}{k,v2}
		    \fmf{dashes}{v2,l}
		    \fmf{dashes}{m,v3}
		    \fmf{scalar}{v3,n}
		    \fmfdot{v1,v3}
	        \end{fmfgraph*}
	    \end{fmffile}
\\\hline
$(\varphi^\dagger\varphi)^2\Phi^2$ &
\begin{fmffile}{e1}
	        \begin{fmfgraph*}(60,25)
 	            \fmfset{arrow_len}{0.25cm}
 	            \fmfset{arrow_ang}{13}
 	            \fmfset{dash_len}{2mm}
	            \fmfbottom{i,l,m}
	            \fmftop{j,k,n}
	            \fmf{scalar}{i,v1}
		    \fmf{scalar}{v1,j}
		    \fmf{dashes}{v1,v2}
		    \fmf{dashes}{v2,v3}
		    \fmf{scalar}{v2,k}
		    \fmf{scalar}{l,v2}
		    \fmf{dashes}{m,v3}
		    \fmf{dashes}{n,v3}
		    \fmfdot{v1,v3}
	        \end{fmfgraph*}
	    \end{fmffile}  & 
\begin{fmffile}{e2}
	        \begin{fmfgraph*}(60,25)
 	            \fmfset{arrow_len}{0.25cm}
 	            \fmfset{arrow_ang}{13}
 	            \fmfset{dash_len}{2mm}
	            \fmfbottom{i,l,m}
	            \fmftop{j,k,n}
	            \fmf{scalar}{i,v1}
		    \fmf{dashes}{v1,j}
		    \fmf{scalar}{v1,v2}
		    \fmf{scalar}{v2,v3}
		    \fmf{scalar}{v2,k}
		    \fmf{scalar}{l,v2}
		    \fmf{dashes}{m,v3}
		    \fmf{scalar}{v3,n}
		    \fmfdot{v1,v3}
	        \end{fmfgraph*}
	    \end{fmffile} \\\hline
$\bar{l}\nu_{R}\tilde{\varphi}\Phi^2$&
\begin{fmffile}{f1}
	        \begin{fmfgraph*}(60,25)
		    \fmfset{arrow_len}{0.25cm}
 	            \fmfset{arrow_ang}{13}
 	            \fmfset{dash_len}{2mm}
	            \fmfbottom{i,l,m}
	            \fmftop{j,n}
	            \fmf{scalar}{i,v1}
		    \fmf{fermion}{v1,j}
		    \fmf{fermion,tension=1}{v2,v1}
		    \fmf{dashes,tension=1}{v2,v3}
		    \fmf{fermion,tension=0.001}{l,v2}
		    \fmf{dashes}{m,v3}
		    \fmf{dashes}{n,v3}
		    \fmfdot{v3}
	        \end{fmfgraph*}
	    \end{fmffile}  & 
\begin{fmffile}{f2}
	        \begin{fmfgraph*}(60,25)
		    \fmfset{arrow_len}{0.25cm}
 	            \fmfset{arrow_ang}{13}
 	            \fmfset{dash_len}{2mm}
	            \fmfbottom{i,l,m}
	            \fmftop{j,n}
	            \fmf{fermion}{i,v1}
		    \fmf{fermion}{v1,j}
		    \fmf{scalar,tension=1}{v2,v1}
		    \fmf{dashes,tension=1}{v2,v3}
		    \fmf{scalar,tension=0.001}{l,v2}
		    \fmf{dashes}{m,v3}
		    \fmf{dashes}{n,v3}          
		    \fmfdot{v3,v2}
	        \end{fmfgraph*}
	    \end{fmffile} \\\hline
\parbox[t][0pt][b]{15pt}{$\bar{l}e{\varphi}\Phi^2$} & & \multirow{3}{*}{\begin{fmffile}{g2}
	        \begin{fmfgraph*}(60,25)
		    \fmfset{arrow_len}{0.25cm}
 	            \fmfset{arrow_ang}{13}
 	            \fmfset{dash_len}{2mm}
	            \fmfbottom{i,l,m}
	            \fmftop{j,n}
	            \fmf{fermion}{i,v1}
		    \fmf{fermion}{v1,j}
		    \fmf{scalar,tension=1}{v2,v1}
		    \fmf{dashes,tension=1}{v2,v3}
		    \fmf{scalar,tension=0.001}{l,v2}
		    \fmf{dashes}{m,v3}
		    \fmf{dashes}{n,v3}
		    \fmfdot{v3,v2}
	        \end{fmfgraph*}
	    \end{fmffile}
} \\
$\bar{q}u\tilde{\varphi}\Phi^2$ &&\\
$\bar{q}d{\varphi}\Phi^2$ & & \\\hline
$\nu^T_R C N_R \Phi^3$ &

\begin{fmffile}{h1}
	        \begin{fmfgraph*}(60,25)
		    \fmfset{arrow_len}{0.25cm}
 	            \fmfset{arrow_ang}{13}
 	            \fmfset{dash_len}{2mm}
	            \fmfleft{i,j}
	            \fmfright{k,l,m}
	            \fmf{fermion}{i,v1}
		    \fmf{fermion}{v1,j}
		    \fmf{dashes}{v1,v2}
		    \fmf{dashes}{v2,k}
		    \fmf{dashes}{v2,l}
		    \fmf{dashes}{v2,m}
	        \end{fmfgraph*}
	    \end{fmffile} & 
\begin{fmffile}{h2}
	        \begin{fmfgraph*}(60,25)
		    \fmfset{arrow_len}{0.25cm}
 	            \fmfset{arrow_ang}{13}
 	            \fmfset{dash_len}{2mm}
	            \fmfbottom{i,l,m}
	            \fmftop{j,n}
	            \fmf{dashes}{i,v1}
		    \fmf{fermion}{v1,j}
		    \fmf{fermion,tension=1}{v2,v1}
		    \fmf{dashes,tension=1}{v2,v3}
		    \fmf{fermion,tension=0.001}{l,v2}
		    \fmf{dashes}{m,v3}
		    \fmf{dashes}{n,v3}
		    \fmfdot{v3}
	        \end{fmfgraph*}
	    \end{fmffile}
\\\hline
$\varphi^\dagger \varphi \nu^T_R C \nu_R \Phi$ &
\begin{fmffile}{i2}
	        \begin{fmfgraph*}(60,25)
		    \fmfset{arrow_len}{0.25cm}
 	            \fmfset{arrow_ang}{13}
 	            \fmfset{dash_len}{2mm}
	            \fmfbottom{i,l,m}
	            \fmftop{j,n}
	            \fmf{scalar}{i,v1}
		    \fmf{scalar}{v1,j}
		    \fmf{dashes}{v2,v1}
		    \fmf{dashes}{v2,v3}
		    \fmf{dashes,tension=0.001}{l,v2}
		    \fmf{fermion}{m,v3}
		    \fmf{fermion}{v3,n}
		    \fmfdot{v1,v2}
	        \end{fmfgraph*}
	    \end{fmffile}
 & 
\begin{fmffile}{i2}
	        \begin{fmfgraph*}(60,25)
		    \fmfset{arrow_len}{0.25cm}
 	            \fmfset{arrow_ang}{13}
 	            \fmfset{dash_len}{2mm}
	            \fmfbottom{i,l,m}
	            \fmftop{j,n}
	            \fmf{scalar}{i,v1}
		    \fmf{scalar}{v1,j}
		    \fmf{dashes}{v2,v1}
		    \fmf{fermion}{v2,v3}
		    \fmf{fermion,tension=0.001}{l,v2}
		    \fmf{dashes}{m,v3}
		    \fmf{fermion}{v3,n}
		    \fmfdot{v1}
	        \end{fmfgraph*}
	    \end{fmffile} 
\\\hline
$\nu^T_R C \gamma^\mu l \varepsilon \partial_\mu \varphi \Phi$ &&
\begin{fmffile}{j1}
	        \begin{fmfgraph*}(60,25)
		    \fmfset{arrow_len}{0.25cm}
 	            \fmfset{arrow_ang}{13}
 	            \fmfset{dash_len}{2mm}
	            \fmfbottom{i,j}
	            \fmftop{k,l}
	            \fmf{fermion}{i,v1}
		    \fmf{fermion}{v1,k}
		    \fmfv{lab=$\gamma_\mu$,lab.dist=0.05w}{v1}
		    \fmfv{lab=$\partial^\mu$,lab.dist=0.05w}{v2}
		    \fmf{boson}{v2,v1}
		    \fmf{dashes}{l,v2}
		    \fmf{scalar}{j,v2}
	        \end{fmfgraph*}
	    \end{fmffile}  \\\hline
\parbox[t][0pt][b]{0pt}{ $\nu^T_{Rp} C \nu_{Rq} \Psi^T C \Psi$} & \multirow{2}{*}{\begin{fmffile}{k1}
	        \begin{fmfgraph*}(60,25)
		    \fmfset{arrow_len}{0.25cm}
 	            \fmfset{arrow_ang}{13}
 	            \fmfset{dash_len}{2mm}
	            \fmfleft{i,j}
	            \fmfright{k,l}
	            \fmf{fermion}{i,v1}
		    \fmf{fermion}{v1,j}
		    \fmf{fermion}{k,v2}
		    \fmf{fermion}{v2,l}
		    \fmf{dashes}{v2,v1}
	        \end{fmfgraph*}
	    \end{fmffile} } & \\
  $\nu^T_{Rp} C \nu_{Rq} \bar{\Psi} C \bar{\Psi}^T$&  & \\\hline
\parbox[t][0pt][b]{0pt}{ $\nu^T_{Rp} C \sigma_{\mu\nu} \nu_{Rq} \Psi^T C \sigma^{\mu\nu} \Psi$} & & \multirow{2}{*}{\begin{fmffile}{l2}
	        \begin{fmfgraph*}(60,25)
		    \fmfset{arrow_len}{0.25cm}
 	            \fmfset{arrow_ang}{13}
 	            \fmfset{dash_len}{2mm}
	            \fmfleft{i,j}
	            \fmfright{k,l}
	            \fmf{fermion}{i,v1}
		    \fmf{fermion}{v1,j}
		    \fmf{fermion}{k,v2}
		    \fmf{fermion}{v2,l}
		    \fmf{dashes}{v2,v1}
	        \end{fmfgraph*}
	    \end{fmffile} } \\
  $\nu^T_{Rp} C \sigma_{\mu\nu} \nu_{Rq} \bar{\Psi} C \sigma^{\mu\nu}\bar{\Psi}^T$&  & \\\hline
  $\bar{\psi} \gamma_{\mu} \psi \bar{\Psi}  \gamma^{\mu} \Psi $ &  
\begin{fmffile}{m1}
	        \begin{fmfgraph*}(60,25)
		    \fmfset{arrow_len}{0.25cm}
 	            \fmfset{arrow_ang}{13}
 	            \fmfset{dash_len}{2mm}
	            \fmfleft{i,j}
	            \fmfright{k,l}
	            \fmf{fermion}{i,v1}
		    \fmf{fermion}{v1,j}
		    \fmf{fermion}{k,v2}
		    \fmf{fermion}{v2,l}
		    \fmf{boson}{v2,v1}
	        \end{fmfgraph*}
	    \end{fmffile}  &
 \\\hline
$i\varphi^\dagger\protect\overleftrightarrow{D}_\mu\varphi\bar\Psi \gamma^\mu \Psi$ &
\begin{fmffile}{n1}
	        \begin{fmfgraph*}(60,25)
		    \fmfset{arrow_len}{0.25cm}
 	            \fmfset{arrow_ang}{13}
 	            \fmfset{dash_len}{2mm}
	            \fmfleft{i,j}
	            \fmfright{k,l}
	            \fmf{scalar}{i,v1}
		    \fmf{scalar}{v1,j}
		    \fmf{fermion}{k,v2}
		    \fmf{fermion}{v2,l}
		    \fmf{boson}{v2,v1}
	        \end{fmfgraph*}
	    \end{fmffile}&\\\hline
$\Lambda^{-2}\varphi^\dagger\varphi V_\mu V^\mu$
&
\begin{fmffile}{o1}
	        \begin{fmfgraph*}(60,25)
		    \fmfset{arrow_len}{0.25cm}
 	            \fmfset{arrow_ang}{13}
 	            \fmfset{dash_len}{2mm}
	            \fmfbottom{i,j}
	            \fmftop{k,l}
	            \fmf{scalar}{i,v1}
		    \fmf{scalar}{v1,k}
		    \fmf{dashes}{v2,v1}
		    \fmf{boson}{l,v2}
		    \fmf{boson}{v2,j}
	        \end{fmfgraph*}
	    \end{fmffile} &\\\hline
\end{tabular}
\caption{The table shows illustrative diagrams that source tree-level generated (PTG)
operators contained in table~\ref{tablesmdm}.
For most cases we present both diagrams generated by an exchange of a mediator that is neutral (second column)
and/or charged (thrid column) under dark and $\nsm$ symmetries. A thick dot stands for a dimensionful 
cubic scalar coupling of the order of $\Lambda$, external lines correspond to $\nsm$ fields while internal ones 
describe propagators of heavy mediators. Dashed, dashed with arrows, solid and wavy lines correspond
to real scalars, complex scalars, fermions and vector bosons, respectively.
}
\label{diag}
\end{center}
\end{table}

\section{Summary}
\label{summary}
\setcounter{equation}{0}
In this paper we have constructed a basis of operators of dim~$\leq 6$
which describe interactions between Dark Matter composed of an Abelian vector, 
chiral fermions and a real scalar with the Standard Model. Our assumptions 
were the following:
\bit
\item
Each component of the dark sector is stable by the virtue of an independent
$\zBB_2$ symmetry,
\item
Each component of the dark sector transforms non-trivially only under
the symmetry which is responsible for its own stability,
\item 
The Standard Model fields
are neutral under any symmetries of the dark sector,
\item
The dark sector contains neutral fermions, $N_{R}$, which are odd under 
$(\zBB_2)_\Phi$ symmetry responsible for stability of
$\Phi$,  the $N_{R}$ has no other quantum numbers, 
\item
The dark sector fields are neutral under any symmetry of the Standard Model,
\eit
The basis consistent with the above assumptions is presented in table~\ref{tablesmdm}. 
where operators redundant under the application of the equations of motion have been eliminated.

We have shown that there exist only two possible operators of dim-4 that are consistent with
our assumptions: the Higgs portal $\varphi^\dagger\varphi \Phi^2$ and 
Yukawa interactions, $\nu^T_{R} C N_{R} \Phi$.

\section*{Acknowledgments}
BG is partially supported by the National Science Centre (Poland) under research project, 
decision no DEC-2011/01/B/ST2/00438.

\appendix
\section{Conventions and definitions}
\label{form}
\setcounter{equation}{0}
In this appendix we collect useful formulae and specify conventions adopted in the main text.
$\tilde\varphi$ is defined as $\tilde{\varphi}_i\equiv\varepsilon_{ij}(\varphi^{j})^*$. 
Tensors $\varepsilon_{ij}$ and $\varepsilon_{\mu\nu\rho\sigma}$ are totally 
antisymmetric with $\varepsilon_{12}=+1$, $\varepsilon_{0123}=+1$.
Dual tensor to $X_{\mu\nu}$ is defined as 
$\tilde{X}_{\mu\nu}=\frac{1}{2}\varepsilon_{\mu\nu\rho\sigma} X^{\rho\sigma}$. 
Symbol $(\sim)$ over ${X}$ denotes $X$ or $\tilde{X}$.
Metric signature $(+,-,-,-)$ is chosen.

Sign convention for covariant derivative is exemplified by
\begin{equation}
 (D_\mu q)^{\alpha j} = \left[(\partial_\mu + i g' Y_q B_\mu)\delta^{\alpha\beta}\delta^{jk} + 
ig_s T^{A\alpha\beta}G^{A}_\mu\delta^{jk}
 + i g S^{Ijk}W^I_\mu\delta^{\alpha\beta}\right]q^{\beta k},
\end{equation}
where $T^A=\frac{1}{2}\lambda^A$ are SU(3) generators with Gell-Mann matrices $\lambda^A$ 
and $S^I=\frac{1}{2}\tau^I$ are SU(2) generators with Pauli
matrices $\tau^I$. It is useful to define Hermitian derivative term
\begin{equation}
 i\varphi^\dagger \overleftrightarrow{D}_\mu \varphi \equiv i \varphi^\dagger D_\mu \varphi - i (D_\mu \varphi)^\dagger \varphi.
\end{equation}
Gauge field strength tensors and their covariant derivatives are 
\begin{equation}
\begin{split}
&G^A_{\mu\nu}=\partial_\mu G^A_\nu - \partial_\nu G^A_\mu - g_s f^{ABC}  G^B_\mu  G^C_\nu,\;\;\;\;
(D_\rho G_{\mu\nu})^A = \partial_\rho G^A_{\mu\nu} - g_s f^{ABC} G^B_\rho  G^C_{\mu\nu} \\
&W^I_{\mu\nu}=\partial_\mu W^I_\nu - \partial_\nu W^i_\mu - g \varepsilon^{IJK}  W^J_\mu  W^K_\nu,\;\;\; 
(D_\rho W_{\mu\nu})^I = \partial_\rho W^I_{\mu\nu} - g \varepsilon^{IJK} W^J_\rho  W^K_{\mu\nu} \\
&B_{\mu\nu}=\partial_\mu B_\nu - \partial_\nu B_\mu,\;\;\;\;\;\;\;\;\;\;\;\;\;\;\;\;\;\;\;\;\;\;\;\;\;\;\;\;\;\;\;\;\;
D_\rho B_{\mu\nu}= \partial_\rho B_{\mu\nu}\\
\end{split}
\end{equation}

\section{Mass generation for Abelian vector bosons}
\label{vbm}
\setcounter{equation}{0}
In this appendix we review possible mechanisms of Abelian vector-boson mass generation.
A massive vector field can be described by the Proca Lagrangian (\ref{proca}), 
since the mass term spoils the gauge invariance therefore renormalizabilty of this theory is not apparent.
\begin{equation}
 \mathcal{L}_P=-\frac{1}{4}V_{\mu\nu}V^{\mu\nu} + \frac{1}{2}m_V^2 V_\mu V^\mu \lsp {\rm for} \lsp
 V_{\mu\nu}=\partial_\mu V_\nu - \partial_\nu V_\mu
\label{proca}
\end{equation}

\subsection{The Stuckelberg mechanism}

The Stuckelberg mechanism is a way to restore the gauge symmetry of (\ref{proca})
by introducing a real scalar field $\sigma$ (see e.g. \cite{Ruegg:2003ps}, \cite{vanHees:2003dk})
with appropriate transformation rules:
\begin{equation}
\begin{split}
&\mathcal{L}_S= -\frac{1}{4}V_{\mu\nu}V^{\mu\nu} + \frac{1}{2}(\partial_\mu \sigma - m_V V_\mu)(\partial^\mu \sigma - m_V V^\mu)\\
&V_\mu \rightarrow V'_\mu= V_\mu +\partial_\mu\chi\\
&\sigma \rightarrow \sigma' = \sigma + m_V\chi.
\label{stuckel}
\end{split}
\end{equation}
The field $\sigma$ can be eliminated from the model by choosing $\chi=-\sigma/m_V$; in this gauge 
the Stuckelberg Lagrangian becomes the same as in the 
Proca theory (\ref{proca}). It should be emphasized that when the mass of the vector 
field is generated by the Stuckelberg mechanism (\ref{stuckel}) the gauge invariance 
requires that the vector field appears only as $V_{\mu\nu}$, $\tilde{V}_{\mu\nu}$ 
or $\mathcal{V}_\mu=\partial_\mu \sigma - m_V V_\mu$, both of which have mass dimension 2.

\subsection{The Higgs mechanism}

Another way to make an Abelian vector field massive is the Higgs mechanism. 
It uses complex scalar field $\phi$ that acquires a \vev\ $\langle \phi\rangle=f/\sqrt{2}$
which spontaneously breaks a $\ui$ local symmetry and thus generates a  mass $ m_V = g f $ for the 
associated gauge vector field $ V_\mu $:
\begin{equation}
\begin{split}
 &\mathcal{L}_H= -\frac{1}{4}V_{\mu\nu}V^{\mu\nu} + (D_\mu\phi)^\dagger D^\mu \phi - \lambda\left(\phi^\dagger\phi - \frac{f^2}{2}\right)^2 \lsp {\rm for} \lsp D_\mu \phi = (\partial_\mu - i g V_\mu)\phi\\
&V_\mu \rightarrow V'_\mu - \partial_\mu \chi, \lsp \phi \rightarrow e^{ig\chi} \phi' . 
\label{higgs}
\end{split}
\end{equation}
In contrast to the Stuckelberg mechanism the scalar field cannot be completely eliminated by the gauge transformation. 
We can write the complex scalar field as $\phi=2^{-1/2}(f+h)e^{i\theta/f}$ in terms of which the Lagrangian becomes
\beq
 \mathcal{L}_H = -\frac{1}{4}V_{\mu\nu}V^{\mu\nu} +
 \frac{1}{2}\left| \partial_\mu h+ig(f+h)(V_\mu - \partial_\mu\theta/m_V )\right |^2  
-\frac{\lambda}{2}(4h^2f^2+4h^3 f + h^4).
\eeq
The field $h$ has a mass $m_h=2\sqrt{\lambda}f$~\footnote{It is worth noticing that the field $\theta$ 
could be removed from the Lagrangian choosing unitary gauge $\phi \rightarrow \phi e^{-i\theta/f}$, 
$V_\mu\rightarrow V_\mu+\partial_\mu\theta /f$. In this gauge the interactions of $h$ and $V_\mu$ are: 
$ 2 g^2 f h V_\mu V^\mu, \;\;\; g^2h^2 V_\mu V^\mu,\;\;\;-2\lambda h^3 f,\;\;\; -\lambda/2 h^4$}. 
The Stuckelberg mechanism can be seen as a limit of the Higgs mechanism
when $ f \to \infty,\, g\to 0 $ with $ m_V$ fixed. The scalar field $h$ decouples from 
$\theta$ and $V_\mu$ as its mass tends to infinity $m_h\rightarrow\infty$ as a result of 
an arbitrarily large dimensional parameter ($f$).
Then $\mathcal{L}_H$ becomes identical as the Stuckelberg one.
\begin{equation}
 \mathcal{L}^{\text{lim}}_H=-\frac{1}{4}V_{\mu\nu}V^{\mu\nu} + \frac{1}{2}(\partial_\mu \theta - m_V  V_\mu)(\partial^\mu \theta -  m_V V^\mu)+\mathcal{L}_h,
\end{equation}
where $\mathcal{L}_h$ is Lagrangian of the decoupled field $h$. 

It is worth to see the decoupling also in the effective field theory framework. 
If we assume that $V_\mu$ is a light particle with mass of the order~$v$, which is the SM Higgs vev, 
we can assign $h$ into the heavy sector of the mass scale $\Lambda$.  
Since the scalar mass is $m_h=2\sqrt{\lambda}f  \sim \sqrt{\lambda} v/g$, therefore for $\lambda \sim 1$ 
it is enough to set coupling $g \sim v/\Lambda$.

Gauge invariant quantities containing a vector field in the model with the Higgs mechanism (\ref{higgs}) are 
built from $V_{\mu\nu}$, $\tilde{V}_{\mu\nu}$
and covariant derivatives of complex scalar field $D_\mu\phi$.  It is assumed that $\nu SM$ fields
are singlets under the Higgs U(1) symmetry. Operators built of $V_{\mu\nu}$ and $\tilde{V}_{\mu\nu}$ 
only are the same as in Stuckelberg case. Operators with $\phi$ appear at dimension 3
(or higher), because they 
must contain $\phi^*$ to ensure gauge invariance and at least one covariant derivative 
that contains the vector field.


\end{document}